# Title: Active sorting of orbital angular momentum states of light with cascaded tunable resonators

**Authors:** Shibiao Wei[1-3,†], Stuart K. Earl[2,3,†], Xiao-Cong Yuan[1,*], Shan Shan Kou[2,*], and Jiao Lin[1,3,4,*]

**Affiliations:**

[1]Nanophotonics Research Centre, Shenzhen University & Key Laboratory of Optoelectronic Devices and Systems of Ministry of Education and Guangdong Province, College of Optoelectronic Engineering, Shenzhen University, Shenzhen 518060, China.
[2]Department of Chemistry and Physics, La Trobe Institute for Molecular Science (LIMS), La Trobe University, Victoria 3086, Australia.
[3]School of Engineering, RMIT University, Melbourne, Victoria 3001, Australia.
[4]School of Physics, The University of Melbourne, Tin Alley, Melbourne, Victoria 3010, Australia.

*Correspondence to: jiao.lin@osamember.org, s.kou@latrobe.edu.au, or xcyuan@szu.edu.cn

†These authors contributed equally to this work.

**Abstract**: Light carrying orbital angular momentum (OAM) has been shown to be of use in a disparate range of fields ranging from astronomy to optical trapping, and as a promising new dimension for multiplexing signals in optical communications and data storage. A challenge to many of these applications is a reliable and dynamic method that sorts incident OAM states without altering them. Here we report a wavelength-independent technique capable of dynamically filtering individual OAM states based on the resonant transmission of a tunable optical cavity. The cavity length is piezo-controlled to facilitate dynamic reconfiguration, and the sorting process leaves both the transmitted and reflected signals in their original states for subsequent processing. As a result, we also show that a reconfigurable sorting network can be constructed by cascading such optical resonators to handle multiple OAM states simultaneously. This approach to sorting OAM states is amenable to integration into optical communication networks and has implications in quantum optics, astronomy, optical data storage and optical trapping.

**Main Text:**

The optical orbital angular momentum (OAM) carried by vortex beams has been shown to be of relevance to a broad range of disciplines, from quantum optics (*1, 2*), high density data storage (*3*) and optical trapping (*4*) to astrophysics (*5, 6*) and telecommunications (*7-11*), even in the optical interferometer for the detection of gravitational waves (*12, 13*). Photons entangled via OAM have also been used to demonstrate the violation of generalized Bell inequalities (*14*) and for quantum cryptography (*15*). Despite this broad range of applications, a method to dynamically discriminate a beam of a specific OAM state from others while retaining the initial state is still lacking.

The OAM of light is associated with the helical phase front of a propagating beam, just as spin angular momentum is connected to the circular polarization of light. Laguerre-Gauss (LG) laser modes were the first to be identified as carrying OAM (*16*), although any beams with an azimuthal phase dependence proportional to *exp(ilθ)* has *lℏ* units of OAM per photon, where the integer number *l* is known as the topological charge, and *θ* is the azimuthal coordinate. A beam of the OAM state $|l=0\rangle$ displays a standard Gaussian intensity distribution, while a beam of any other OAM state has an intensity void (optical vortex) in its center due to a phase singularity there arising from the helical phase profile (*17*).

Passing through an optical element that has the opposite helical phase *exp(-ilθ)*, a vortex beam of the topological charge +*l* can be converted into a Gaussian beam ($|l=0\rangle$), which can then be coupled into a single-mode optical fibre. All other co-propagating OAM states will be unable to couple into the fiber due to the singularity at their center, thus forming a simple mode-selection device (*18*). A more intricate conformal transformation has been used to convert the helical phase structure of a vortex beam into a linear phase gradient using either spatial light modulators (SLMs) (*19*) or custom-designed refractive elements (*20*). The resultant linear spread of the beam profile is able to spatially disperse OAM states based on the topological charge. Spatial dispersion of OAM states was also recently reported using a plasmonic metasurface (*3*), in which nanostructured grooves in a metallic film coupled incident OAM states to surface plasmon polaritons on the metal surface, subsequently spatially-routing them to nano-ring slits based on their topological charge.

While spatially dispersing vortex beams through the conversion of the helical phase is an ingenious approach, it destroys the incident OAM states and is therefore more suitable for detecting the topological charge at the endpoint of an optical system. A more general sorting mechanism would be one that sends different OAM states to different output ports *without* altering the original states. This concept has been attempted by a modified Mach-Zehnder interferometer consisting of Dove prisms and spiral phase plates to construct a sorting device to select beams based on their OAM, *l* (*2*) and total angular momentum, *j=s+l*, where *s* is spin angular momentum (*1*). Nevertheless, this method still changes incident OAM states by using spiral phase plates, although the recovery of the original states is relatively easy. In theory, the interferometric method can sort a number of OAM states. However, sorting devices using this approach require (*n*-1) interferometer arms grouped at several stages to sort *n* different OAM states, meaning the complexity and losses increase rapidly with the number of OAM states. A unique combination of the configurations of all interferometer arms needs to be found for a given set of OAM states to be sorted for. The configuration at each stage of this form of interferometer network may then also need to be completely modified with the inclusion of each new OAM state.

In this work, we propose the modular sorting process illustrated in Fig. 1A, where each module must accept a number of input OAM states, output only one and divert all the others unaltered for subsequent processing. Therefore, within the modular architecture, the design of an OAM sorting system may be realized by simply cascading the modules and the inclusion of any new OAM state will be easily achieved by adding an extra module towards the end of the chain. In order for the cascaded arrangement to work, it is important that each module rejects non-output OAM states with very high efficiency so that the last output OAM state would not suffer from a significant intensity drop relative to the first output. Mirrors are reflective optical components that can reflect (reject) incident light at almost 100% efficiency (99.9% readily achieved by modern mirrors over the visible or infrared spectrum) while preserving most of the characteristics of the original light. However, a normal mirror cannot differentiate OAM states and acts like a barrier to all incident photons. Surprisingly, by adding another mirror (the second barrier, see the inset of Fig. 1A), a particular OAM state can tunnel through the double barriers formed by the pair of mirrors in a fashion similar to a resonant tunneling diode. The space between the two mirrors can be considered an optical cavity that resonates with the selected OAM state, resulting in the tunneling of the state through the two mirrors. This configuration is, in fact, a *Fabry-Pérot* (FP) cavity (*21*), a ubiquitous, high-finesse optical resonator appearing in numerous applications ranging from large-scale interferometers (*22*) and resonant mode cleaning (*23*) to astronomy (*24*) and spectroscopy (*25*). The transmission of an FP cavity can approach 100% with extremely high resolution (*26*), and the resonant properties can be manipulated by altering the cavity length, incident angle and reflectivity of the mirrors (*27, 28*). Practically, most FP cavities are formed by two curved mirrors instead of parallel plane mirrors to reduce walk-off and to simplify alignment of the incident beam. A propagating OAM state acquires a phase shift relative to a plane wave as it is focused by the curved mirrors. The accumulated phase shift varies with the helical wavefront and determines whether the state can form a standing wave in the cavity, i.e. at resonance. An *implicit* relation between the cavity length and the topological charge of a resonant OAM state can be found (Supplementary Information Section A):

$$D = \frac{\lambda}{2}\left[q + \left(|\ell| + 1\right)\frac{\varphi}{\pi}\right], \qquad (1)$$

where $D$ is the cavity length; $\varphi = \arccos\left(\pm\sqrt{(1 - D/R_1)(1 - D/R_2)}\right)$ is the accumulated Gouy phase shift experienced by the light traveling from one end of the cavity to the other, determined by the mirror curvatures ($R_1$ & $R_2$); $\lambda$ is the wavelength of light and $q$ is an integer. Since the cavity is sensitive to even subwavelength (e.g. a few hundred nanometers) movement, the transcendental equation can be approximated by an *explicit* relation between the small variation $\Delta$ of cavity length and the resonant topological charge (Supplementary Information Section B):

$$\Delta = \frac{\lambda}{2}\left[q + \left(|\ell|\right)\frac{\varphi}{\pi}\right], \qquad (2)$$

Hence, one can resonantly match a specific OAM state to an FP cavity by tuning its cavity length, providing an OAM filter that allows the transmission of the selected state and rejects all others. The resonant filter then serves as the key component in the constituent module of an OAM sorting system.

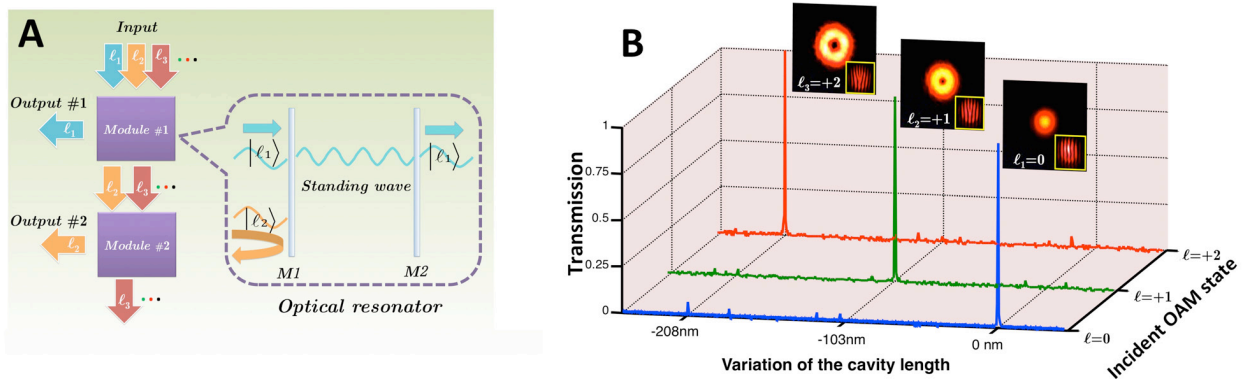

*Figure 1: **(A)** Conceptual diagram of a desired OAM sorting system based on a cascaded modular design. For example, three different OAM states ($l_1$, $l_2$ and $l_3$) enter the system. First module outputs one state ($l_1$) while rejecting and redirecting the rest ($l_2$ and $l_3$) to the next module where the state ($l_2$) is separated from the other ($l_3$) and exits the sorter. The main component of a module here is an optical cavity/resonator tuned to a specific incident OAM state. M1 & M2 are highly reflective mirrors forming a Fabry-Pérot cavity. The resonant OAM state will experience high transmission through the cavity due to resonant tunneling while the non-resonant states experience destructive interference within the cavity and are therefore completely reflected. **(B)** Measured transmission of various OAM states as a function of the change in length of a Fabry-Pérot cavity. Experimental details are supplied in the Supplementary Information section C. The shift of the transmission peak, as measured using a photodiode, shows a clear dependency on $|l|$. Some small additional peaks are parasitic cavity modes coming from mirror imperfections, slight misalignment/asymmetry of the incident beam and the impurity within the incident OAM state. Insets: Intensity distribution of the transmission and respective fork interference pattern with a reference plane wave. The number of the branches at the dislocation in the resulting interference pattern enables the identification of the value of the topological charge ($|l|$) of the OAM state, validating our claim that the transmitted beam retains its initial state.*

A scanning FP cavity whose length can be actively tuned by a piezoelectric transducer was selected for the sorting experiment. A liquid-crystal-based spatial light modulator (SLM) was used to produce various incident OAM states from a frequency-stabilized HeNe laser of 633nm (*29*). Figure 1B clearly shows that a sharp transmission peak occurs at different cavity lengths for each OAM state. In other words, the FP acts as a filter, transmitting a specific OAM state for the correctly selected cavity length. At resonance, we also verified that the transmitted vortex beam remains in its original OAM state. These experimental results are corroborated by numerical calculations performed using FINESSE (*30*), an interferometer simulation package. The (absolute) cavity length used in these simulations was calculated using the change in cavity length relative to that of a Gaussian beam using the $l$-dependent Gouy phase accumulated by a vortex beam within the FP (see Supplementary Information Section A).

**Separation of two co-propagating OAM states**

As an essential part of an OAM sorter, the FP cavity must be able to isolate a specific OAM state from a number of co-propagating states. To create a laser beam carrying two distinct charges using a single SLM an iterative algorithm (*31*) was used to create the appropriate phase pattern, which was then displayed on the SLM to convert the incident Gaussian beam into the combination of vortex beams used.

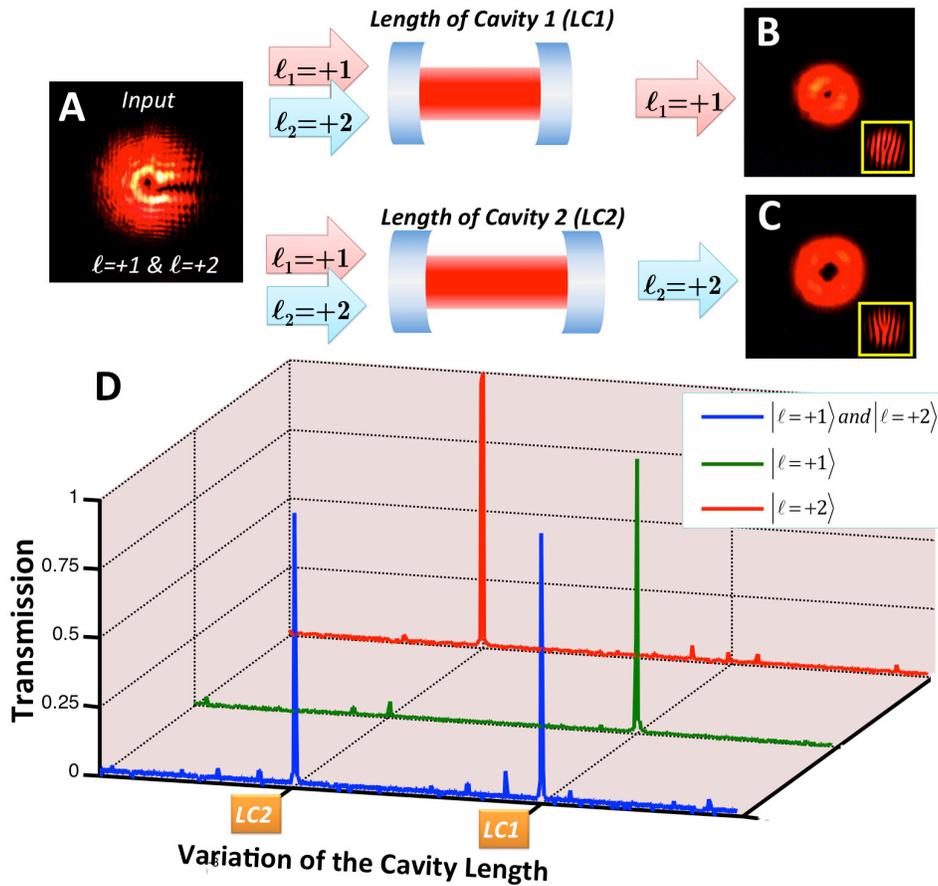

*Figure 2: (A) Incident and (B),(C) transmitted intensity distributions of the superposition of two OAM states before & after the FP cavity on resonance. The topological charge of the transmitted OAM state is quantified by the forked interference pattern. (D) Transmission of the superposition of two l=1&2 states as a function of cavity length (blue line), accompanied by reference spectra of two single OAM states (green, l=+1 and red l=+2). An additional source of the smaller peaks in this experiment emerges from the inherent errors in the iterative algorithm used to generate the superimposed states.*

Figure 2 shows that the FP cavity disperses the input beam, transmitting each constituent OAM state at different cavity lengths. The two distinctive transmission peaks occur at the same cavity lengths as those of the reference beams with a single OAM state.

**Modular design of an OAM sorter**

To those off-resonant OAM states, the FP cavity is simply two mirrors with very high reflectances. The high reflectivity, coupled with the fact that the cavity preserves all incident OAM states (in both transmission and reflection), enables the implementation of a modular design in which multiple cavities are cascaded, allowing the simultaneous sorting of multiple OAM states. To demonstrate this, a second FP cavity was added to construct an OAM sorter with two output ports, as shown in Figure 3. An optical circulator is used to direct the reflected states from the first cavity to the second. We show that it is possible to filter separate OAM states using cascaded cavities, enabling the simultaneous detection of multiple co-propagating OAM states within a single beam. The path taken by a particular OAM state within the system can be manipulated in real time by altering the cavity lengths, permitting the dynamic redirection of an OAM state to any output port. Since the resonant cavity length only depends on the modulus of the topological charge, as indicated by Eq. 1, two OAM states with topological charges of the opposite sign are degenerate, and could not be differentiated by a single FP cavity. One can choose to either use only positive (or negative) topological charges in the system or add a spiral phase plate and a second FP cavity to the module (Supplementary Information Section E).

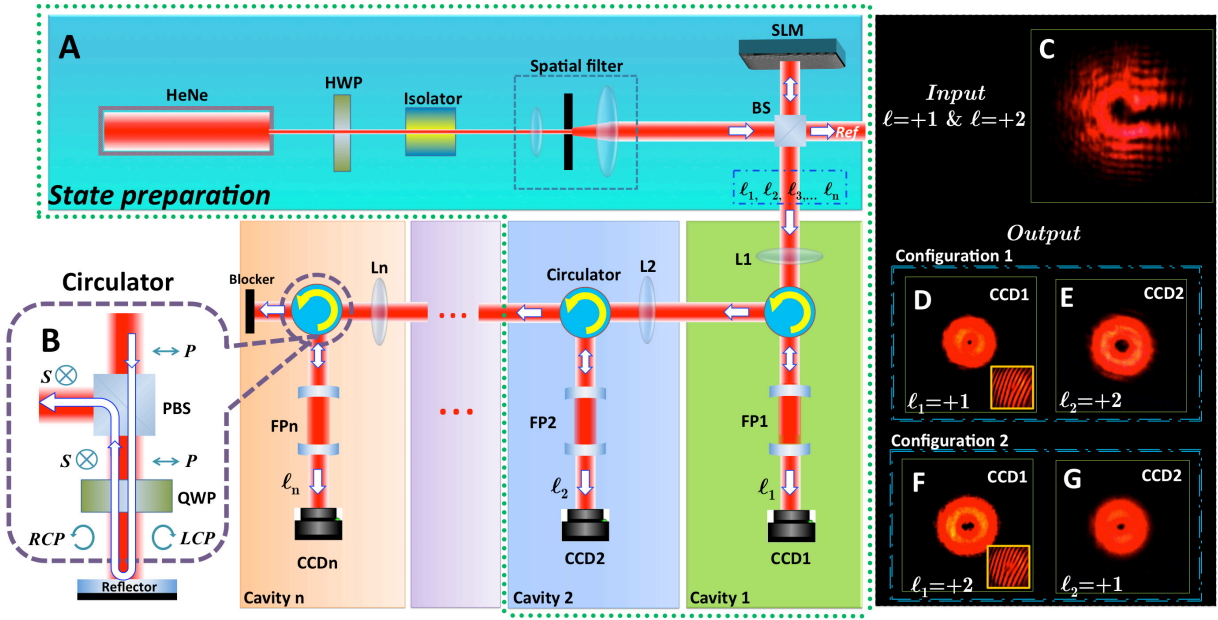

*Figure 3:(A) Experimental setup of the OAM sorter. A half-wave plate (HWP) is used to rotate the linear polarization axis of the laser beam prior to spatial filtering to facilitate alignment to the optical axis of the SLM. An isolator is used to prevent unwanted reflected light from entering the frequency-stabilized laser cavity. The superimposed OAM states are prepared by the SLM loaded with specially designed patterns. A reference beam is generated from one of the ports of a non-polarizing beam-splitter (BS) for interference with the output OAM states in the latter part of the experiment. A lens (L1, 250 mm focal length) is used to mode match the incident Gaussian beam to the first cavity (FP1) by transforming the beam parameters to match the cavity mode. The transmitted signal from each FP cavity was monitored using either a photodiode or a CCD camera. While only two FP cavities (enclosed in the dashed box) were used in this experiment, the high sensitivity and signal-to-noise ratio of the cavities allow a number of them to be cascaded to maximise the number of OAM states being sorted. (B) Optical circulator (32) consisting of a polarizing beam-splitter (PBS) and a quarter-wave plate (QWP). The incident light is p-polarized (linear polarization). Light reflected from the FP cavity passes the QWP twice, transformed into the orthogonal s-polarization state and therefore exits the PBS from a different port. RCP: right-handed circular polarization; LCP: left-handed circular polarization. (C)-(G) Intensity distributions and forked interference patterns when the pair of FP cavities are each transmitting a different OAM state. These images were taken at the two output ports when both cavities were held on resonance for one of the two OAM states. To capture these images, the first cavity was held on resonance to transmit $|l=+1\rangle$ and the second to transmit $|l=+2\rangle$ (and vice versa). We have also shown that the output OAM states can be swapped between the two ports dynamically by varying the cavity lengths (Configuration 1 vs Configuration 2).*

## Simultaneous sorting of OAM and wavelength

FP cavities are regularly used to differentiate finely-spaced laser lines. This ability complements our approach to the sorting of OAM states, permitting simultaneous differentiation of optical signals based on both wavelength and the OAM state within the same device. Here we demonstrate simultaneous sorting of wavelengths and OAM states using a single FP cavity.

The frequency-stabilized HeNe laser used in the previous sections of this paper was replaced with another HeNe laser that is known to lase on 3 longitudinal modes (Melles Griot 05-LHR-151, mode spacing 438 MHz, producing 3 laser lines with a separation of approximately 0.000584 nm) centred on 632.816 nm in air. The remainder of the experimental setup was kept unchanged as in Figure 1. The three laser lines, which would be indistinguishable using a diffraction grating, are clearly visible in Figure 4 due to the superior resolution of the FP cavity.

An input superposition of two OAM states (*+1* and *+2*) across the three wavelengths of the laser were directed through the FP cavity. The results of this experiment are shown in Figure 4 (red line), with the additional results from single OAM states for reference. The six transmission peaks account for all possible combinations of the two OAM states and three wavelengths. These results show that the OAM sorter can be used to separate wavelengths simultaneously without requiring additional components.

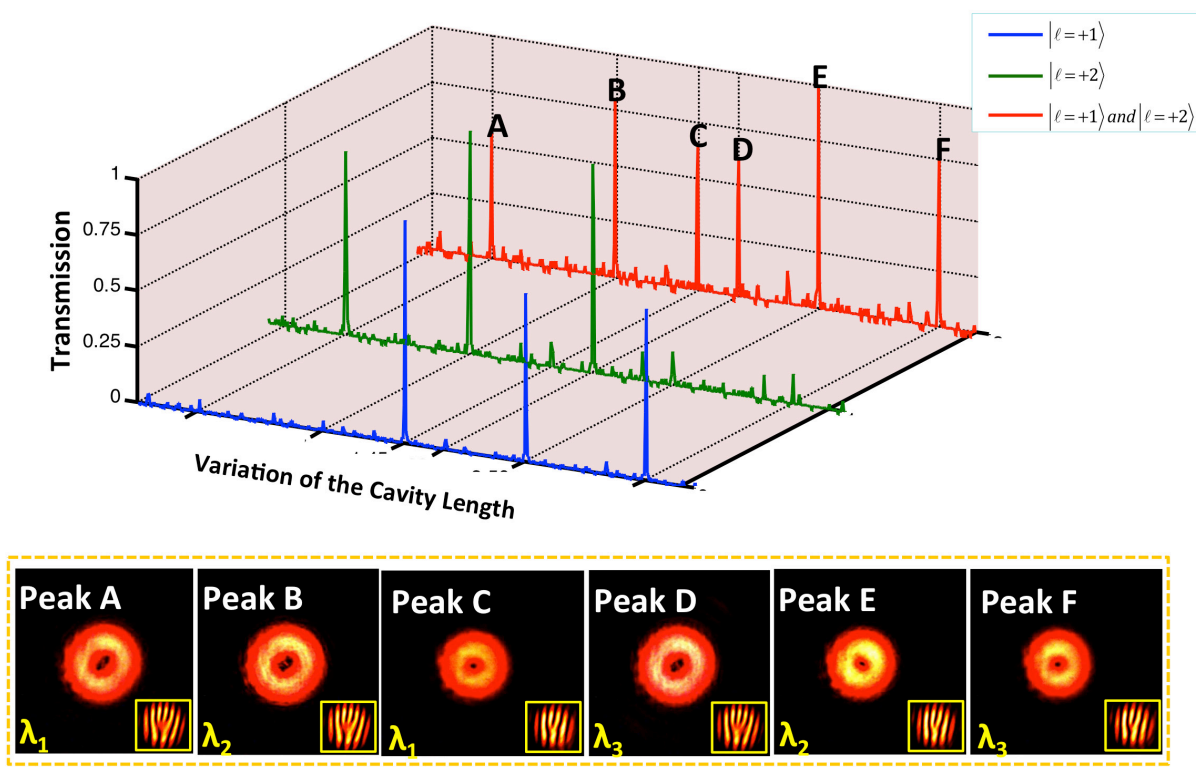

*Figure 4: Simultaneous OAM and wavelength multiplexing in an incident beam. The blue and green line are single OAM states generated from the same laser (with three laser lines) for reference, while the red line represents an incident beam of both OAM states at the three laser lines. The peaks of the mixed OAM states match those of individual OAM states. Insets: CCD images of each of the six resonant transmissions.*

## Discussion:

We have demonstrated a reconfigurable OAM sorter with a modular design, which can sort multiple OAM states simultaneously while preserving the original states. The system can change its sorting function on the fly (with the theoretical switching time down to a few nanoseconds) without physically rearranging the constituent optical elements. The basic building element of the OAM sorter is an optical cavity and thus the sorter inherits a number of merits from using a resonance-based approach, such as high selectivity and high efficiency. The high resolution of FP cavities provides a high sensitivity sorting method that can identify finely-spaced wavelengths, even at low powers, while the high reflectivity experienced by non-resonant beams facilitates the cascading of multiple cavities. The cross-talk among OAM states is suppressed due to the resonant basis of this approach. The transmission of the resonant OAM states is experimentally measured to be around 20%, which is very close to the maximum efficiency (21.7%) of this particular off-the-shelf FP cavity we used in the experiment for an input mode-matched Gaussian beam. The transmission efficiency can be significantly improved (e.g. over 99% transmission efficiency reported experimentally by using mirrors with higher reflectance in the cavity and moving to longer wavelengths such as 1064nm (Supplementary Information Section F) (*33*).

In conjunction with wavelength and polarization, mode-division multiplexing (MDM) using optical OAM states has been investigated due to the infinite number of orthogonal states potentially available for increasing the transmission capacity of a network. Already, OAM-enabled optical networks have demonstrated greater than Terabit per second capacities (*7, 8, 34*). While increases in the data-carrying capacity of single mode fibers has outstripped demand over the previous three decades, our ever-expanding demand for more data is predicted to overcome current capacities within a decade (*35*). The dynamic nature of the OAM sorter offers a simple avenue to construct reconfigurable optical networks to address this shortfall. The mature nature of optical cavity technology, the high quality available materials and the ability to reduce the cavity dimensions down to on-chip and in-fiber cavities all indicate a broad impact for this approach. Our OAM sorting technique may also be of interest to imaging, as hyperspectral imaging CCD arrays now integrate fabry-perot cavities on a single-pixel level (*36*), and quantum optics researchers may find interesting implications in these findings for cavity QED experiments (*37, 38*).

**Acknowledgments:** The authors would like to thank Professor Baohua Jia (Swinburne University of Technology) and Professor Ann Roberts (The University of Melbourne) for the loan of equipment.

**Any Additional Author notes:** JL conceived of the experiment. SW and SKE performed the experiment. SW, SKE, SSK and JL acquired, analyzed and interpreted the data. SKE drafted the manuscript with inputs from all other authors. JL, SSK and XY supervised the project and revised the manuscript critically for important intellectual content.

1. J. Leach *et al.*, Interferometric methods to measure orbital and spin, or the total angular momentum of a single photon. *Physical review letters* **92**, 013601 (2004).
2. J. Leach, M. J. Padgett, S. M. Barnett, S. Franke-Arnold, J. Courtial, Measuring the orbital angular momentum of a single photon. *Physical review letters* **88**, 257901 (2002).

3. H. Ren, X. Li, Q. Zhang, M. Gu, On-chip noninterference angular momentum multiplexing of broadband light. *Science* **352**, 805-809 (2016).
4. H. He, M. Friese, N. Heckenberg, H. Rubinsztein-Dunlop, Direct observation of transfer of angular momentum to absorptive particles from a laser beam with a phase singularity. *Physical Review Letters* **75**, 826 (1995).
5. M. P. Lavery, F. C. Speirits, S. M. Barnett, M. J. Padgett, Detection of a spinning object using light's orbital angular momentum. *Science* **341**, 537-540 (2013).
6. G. C. Berkhout, M. W. Beijersbergen, Method for probing the orbital angular momentum of optical vortices in electromagnetic waves from astronomical objects. *Physical review letters* **101**, 100801 (2008).
7. N. Bozinovic *et al.*, Terabit-scale orbital angular momentum mode division multiplexing in fibers. *Science* **340**, 1545-1548 (2013).
8. T. Lei *et al.*, Massive individual orbital angular momentum channels for multiplexing enabled by Dammann gratings. *Light: Science & Applications* **4**, e257 (2015).
9. Y. Yan *et al.*, High-capacity millimetre-wave communications with orbital angular momentum multiplexing. *Nature communications* **5**, (2014).
10. G. Gibson *et al.*, Free-space information transfer using light beams carrying orbital angular momentum. *Optics Express* **12**, 5448-5456 (2004).
11. J. Lin, X.-C. Yuan, S. Tao, R. Burge, Multiplexing free-space optical signals using superimposed collinear orbital angular momentum states. *Applied optics* **46**, 4680-4685 (2007).
12. M. Bondarescu, K. S. Thorne, New family of light beams and mirror shapes for future LIGO interferometers. *Physical Review D* **74**, 082003 (2006).
13. B. Mours, E. Tournefier, J.-Y. Vinet, Thermal noise reduction in interferometric gravitational wave antennas: using high order TEM modes. *Classical and Quantum Gravity* **23**, 5777 (2006).
14. A. C. Dada, J. Leach, G. S. Buller, M. J. Padgett, E. Andersson, Experimental high-dimensional two-photon entanglement and violations of generalized Bell inequalities. *Nature Physics* **7**, 677-680 (2011).
15. S. Gröblacher, T. Jennewein, A. Vaziri, G. Weihs, A. Zeilinger, Experimental quantum cryptography with qutrits. *New Journal of Physics* **8**, 75 (2006).
16. L. Allen, M. W. Beijersbergen, R. Spreeuw, J. Woerdman, Orbital angular momentum of light and the transformation of Laguerre-Gaussian laser modes. *Physical Review A* **45**, 8185 (1992).
17. A. M. Yao, M. J. Padgett, Orbital angular momentum: origins, behavior and applications. *Advances in Optics and Photonics* **3**, 161-204 (2011).
18. A. Mair, A. Vaziri, G. Weihs, A. Zeilinger, Entanglement of the orbital angular momentum states of photons. *Nature* **412**, 313-316 (2001).
19. G. C. Berkhout, M. P. Lavery, J. Courtial, M. W. Beijersbergen, M. J. Padgett, Efficient sorting of orbital angular momentum states of light. *Physical review letters* **105**, 153601 (2010).
20. M. P. Lavery *et al.*, Refractive elements for the measurement of the orbital angular momentum of a single photon. *Optics express* **20**, 2110-2115 (2012).
21. C. Fabry, A. Perot, Theorie et applications d'une nouvelle methode de spectroscopie interferentielle. *Ann. Chim. Phys* **16**, 115 (1899).
22. W. E. Althouse, M. E. Zucker, LIGO: The laser interferometer gravitational-wave observatory. *Science* **256**, 325 (1992).
23. A. Rüdiger *et al.*, A mode selector to suppress fluctuations in laser beam geometry. *Journal of Modern Optics* **28**, 641-658 (1981).
24. T. De Graauw *et al.*, Observing with the ISO Short-Wavelength Spectrometer. *Astronomy and Astrophysics* **315**, L49-L54 (1996).
25. P. Jacquinot, New developments in interference spectroscopy. *Reports on Progress in Physics* **23**, 267 (1960).
26. A. E. Siegmann. (University Science Books, 1986).
27. H. Kogelnik, T. Li, Laser beams and resonators. *Applied optics* **5**, 1550-1567 (1966).
28. L. Pedrotti, SJ Leno M. *Pedrotti Leno S. Pedrotti, Frank "Introduction to Optics third edition" p17-25*, (2007).
29. D. L. Andrews, *Structured light and its applications: An introduction to phase-structured beams and nanoscale optical forces*. (Academic Press, 2011).
30. A. Freise, D. Brown, C. Bond, Finesse, Frequency domain INterferomEter Simulation SoftwarE. *arXiv preprint arXiv:1306.2973*, (2013).

31. J. Lin, X.-C. Yuan, S. Tao, R. Burge, Collinear superposition of multiple helical beams generated by a single azimuthally modulated phase-only element. *Optics letters* **30**, 3266-3268 (2005).
32. J.-M. Liu, *Photonic devices*. (Cambridge University Press, 2009).
33. H. Sekiguchi *et al.*, Ultralow-loss mirror of the parts-in-10 6 level at 1064 nm. *Optics letters* **20**, 530-532 (1995).
34. J. Wang *et al.*, Terabit free-space data transmission employing orbital angular momentum multiplexing. *Nature Photonics* **6**, 488-496 (2012).
35. R.-J. Essiambre, R. W. Tkach, Capacity trends and limits of optical communication networks. *Proceedings of the IEEE* **100**, 1035-1055 (2012).
36. P. Agrawal *et al.*, Characterization of VNIR hyperspectral sensors with monolithically integrated optical filters. *Electronic Imaging* **2016**, 1-7 (2016).
37. C. Hood, M. Chapman, T. Lynn, H. Kimble, Real-time cavity QED with single atoms. *Physical review letters* **80**, 4157 (1998).
38. C. Hood, T. Lynn, A. Doherty, A. Parkins, H. Kimble, The atom-cavity microscope: Single atoms bound in orbit by single photons. *Science* **287**, 1447-1453 (2000).

# Supplementary Materials:

The Supplementary Materials include numerical calculations supporting our findings, experimental details and a discussion on the accumulated Gouy phase of a beam within an optical cavity.

## Materials and Methods:

### Section A—The enabling mechanism of the OAM sensitivity of an optical cavity: the accumulated Gouy phase shift

The mechanism behind the OAM-sorting ability of a Fabry-Pérot (FP) cavity is the Gouy phase shift that the resonant beam accumulates within the cavity. This phenomenon, in which a propagating wave acquires a phase shift relative to a (theoretical) plane wave as it is focused by an optical system, was first observed in 1890 (*1*).

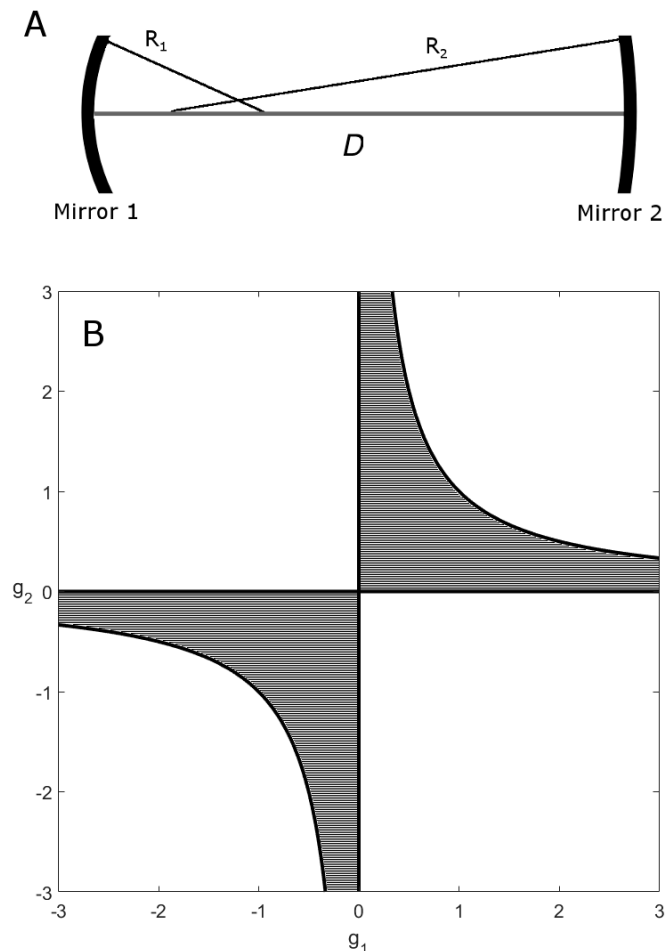

*Figure S5: (A) Schematic of a generic two-mirror Fabry-Pérot cavity comprised of two concave mirrors. Radii of curvature are marked on the figure; (B): Stability diagram of the optical cavity.*

*The shaded regions indicate the stable cavity regions. The hyperbola bounding the shaded regions represent $g_1g_2 = 1$.*

Figure S1A shows a schematic of a generic two-mirror optical cavity comprised of concave mirrors with different radii of curvature. The stability of such a cavity can be calculated by ray transfer analysis (*2*). The result of this analysis commonly introduces the *g* factors of a cavity, defined as

$$g_{1,2} = 1 - D/R_{1,2}, \tag{S1}$$

$g_{1,2} = 1 - D/R_{1,2}$ where $R_{1,2}$ is the radius of curvature of the specified mirror and *D* is the mirror separation, such that a cavity is stable provided that

$$0 \leq g_1 g_2 \leq 1. \tag{S2}$$

A stable cavity is one in which a paraxial ray injected into the resonator does not escape, but rather, remains confined near the longitudinal axis of the cavity. In contrast, an unstable cavity will allow light, even on-axis light, to eventually escape. Figure S1B shows the combinations of the *g* factors that result in a stable cavity as the shaded region. The commercial scanning fabry-perot used in our experiment was confocal ($D = R_1 = R_2$), placing it at the origin (0,0) in the above diagram. The advantage of a confocal FP cavity is that all longitudinal and transverse modes arise at the same cavity length. Since this experiment relied on the cavity to disperse the input spatial modes, the cavity length was extended ($R_1 = R_2 = R < D < 2R$) so that this was no longer the case. The resultant cavity sits in the lower left quadrant of the stability diagram due to the convention of concave mirrors having a positive radius of curvature in this context.

In the study of the detailed field distributions in the FP cavity, we chose Laguerre-Gauss (LG) modes, a family of solutions to the paraxial wave equation in cylindrical coordinates, which, as discussed, were previously shown to carry OAM (*3*). The general solutions are of the form

$$u_{pl}(r,\theta,z) = \frac{C}{\sqrt{1+z^2/z_R^2}} \left(\frac{r\sqrt{2}}{w(z)}\right)^l L_p^l\left(\frac{2r^2}{w(z)^2}\right) \times exp\left(\frac{-r^2}{w(z)^2}\right) exp\left(\frac{-ikr^2 z}{2(z^2+z_R^2)}\right) \times exp(-il\theta) exp\big(i(2p+|l|+1)\psi(z)\big). \tag{S3}$$

In this equation $u_{pl}$ is the eigensolution to the paraxial wave equation; *C* is a normalization constant; $L_p^l$ is the Laguerre-Gauss equation of order (*p,l*); *p* and *l* are the radial and azimuthal mode indexes, respectively; *w(z)* is the standard definition of the beam waist; *r*, *θ* and *z* are the radial, azimuthal and longitudinal coordinates, respectively and $z_R$ is the Raleigh range of the beam, defined as

$$z_R = \frac{\pi w_0^2}{\lambda}. \tag{S4}$$

$w_0$ is the standard definition of the beam waist. The Raleigh range gives the distance the beam needs to propagate from the waist for the width of the beam to increase by √2, which corresponds to an on-axis intensity of half the peak intensity at that point.

The term $\psi(z) = tan^{-1}(z/z_R)$ represents the Gouy phase shift as the beam propagates through the region around its focal point. The above equation shows that for higher-order *LG* modes there is an additional Gouy phase shift relative to a Gaussian beam due to the additional transverse structure of the beam. It is this additional phase shift that enables a resonant cavity to sort vortex beams of different OAM states.

A resonant mode of a cavity was defined by Kogelnik as a self-consistent field configuration (*2*); if a mode can be represented as a wave traveling back and forth between the mirrors, the beam parameters must by necessity be unchanged after one return trip around the cavity. This implies that in an optical cavity, a resonance may only occur when the phase shift from one mirror to the other is a multiple of π (*2*).

The total phase accumulated by an *LG* beam travelling from one side of a stable optical cavity to the other (ie mirror 1 at $z_1$ and mirror 2 at $z_2$) can be written as

$$\emptyset(z_2 - z_1) = kD - (2p + |l| + 1)[\psi(z_2) - \psi(z_1)] \tag{S5}$$

The first term on the right hand side represents the phase advance due to the light of propagation constant $k$ travelling a distance $D$, while the second represents the accumulated Gouy phase within the cavity. This second term was shown by Siegman to be reliant only on the cavity parameters (4), ie

$$\psi(z_2) - \psi(z_1) = \varphi = arccos(\pm\sqrt{g_1 g_2}), \tag{S6}$$

where the $g$ factors are as defined previously. The choice of sign in the definition of the Gouy phase depends on where in the stability diagram (Figure S1B) the cavity in question sits. The '+' sign applies to the upper right quadrant, while the '-' sign applies to the lower left quadrant.

Coupled with the previous equation, and the knowledge that the condition for a standing wave is that the total round trip (total distance travelled = 2D) phase must be an integer multiple of $2\pi$, to be resonant with the cavity a mode must satisfy the equation

$$\frac{\omega 2D}{c} - 2(2p + |l| + 1)\varphi = 2q\pi, \quad q \text{ is an integer.} \tag{S7}$$

Here, the substitution $k = \frac{\omega}{c}$ was used to facilitate the derivation of the resultant resonant frequencies of the longitudinal-plus-transverse modes of the cavity, ie

$$\omega = \omega_{qpl} = \frac{\pi c}{D}\left(q + \frac{(2p+|l|+1)}{\pi}\varphi\right), \tag{S8}$$

where the factors of 2 have been cancelled. This equation tells us that for a cavity of fixed length $D$, different $LG$ modes will resonate with the fixed cavity at slightly different frequencies due to the influence of the Gouy phase.

For a dynamic cavity such as the scanning FP cavity used in this experiment, it is possible to rearrange the previous equation, as $D$ is variable while the frequency is fixed. The result is that for an $LG_p^l$ mode of wavelength λ within a cavity, the resonant cavity length, $D$, is (4)

$$D = \frac{\lambda}{2}\left\{q + (2p + |l| + 1)\frac{\varphi}{\pi}\right\}, \tag{S9}$$

Here, $\lambda = 2\pi c/\omega$ was used to simplify the resulting equation. Since we are interested in the sorting of OAM (the azimuthal mode index $l$), the radial mode index $p$ is irrelevant in this case and thus set to zero to further simplify the equation:

$$D = \frac{\lambda}{2}\left\{q + (|l| + 1)\frac{\varphi}{\pi}\right\}. \tag{S10}$$

This equation shows that due to the accumulated Gouy phase shift experienced by a beam when travelling from one mirror to the other, the resonant length of the cavity is dependent on the values of the azimuthal mode index (OAM state). This dependence is the mechanism that enables the FP cavity to differentiate vortex beams of different OAM states. As the cavity length is varied, different OAM states resonate within the cavity (and are transmitted) while all others experience destructive interference (and are therefore reflected).

### Section B—The linear relation between the variation of the cavity length and the topological charge of the resonant OAM state.

We begin by expanding equation (S10) (Equation (1) in the main text) and define D within this section as the resonant length of the cavity for a Gaussian beam ($p = l = 0$), ie

$$D = \frac{\lambda}{2}\left\{q + \frac{\varphi}{\pi}\right\} = \frac{\lambda}{2}\left\{q + \frac{arccos(\pm\sqrt{g_1 g_2})}{\pi}\right\} = \frac{\lambda}{2}\left\{q + \frac{arccos\left(\pm\sqrt{(1-D/R_1)(1-D/R_2)}\right)}{\pi}\right\} \tag{S11}$$

By fixing $D$ as the initial cavity length, the change in cavity length for other $LG$ modes becomes a correction (Δ) to that length. The equation for cavity length as a function of radial and azimuthal mode index then becomes

$$D + \Delta = \frac{\lambda}{2}\left\{q' + (2p + |l| + 1)\frac{arccos\left(\pm\sqrt{\left(1 - (D+\Delta)/R_1\right)\left(1 - (D+\Delta)/R_2\right)}\right)}{\pi}\right\}$$

$$\tag{S12}$$

We can substitute the definition for D using equation (S11), and, after collecting like terms, arrive at

$$\Delta = \frac{\lambda}{2}\left\{(q' - q) + \frac{2p+|l|}{\pi} arccos\left(\pm\sqrt{\left(1 - D/R_1\right)\left(1 - D/R_2\right)}\right)\right\}. \quad (S13)$$

Here an approximation is made that $(D+\Delta)/R_{1,2} \sim D/R_{1,2}$. This is a valid assumption due to the relative orders of magnitude of D and R compared to that of $\Delta$; as $\Delta$ is on the order of a fraction of a wavelength, it is approximately five orders of magnitude smaller than both D and R.

By taking $q'' = q' - q; p = 0,$ and $\varphi = arccos(\pm\sqrt{g_1 g_2})$, we arrive at the linear relation:

$$\Delta = \frac{\lambda}{2}\left\{q'' + |l|\frac{\varphi}{\pi}\right\}. \quad (S14)$$

In relation to the experiment under discussion here there are two further simplifications that can be made. The two cavity mirrors are identical, having radii of curvature, R, equal to 50mm each, simplifying the argument of the final term. A consideration of the experimental technique requires that an additional assumption is made. Due to having measured the change in resonant cavity length to that of the same Gaussian transmission peak each time, the implicit assumption has been made that $q' = q$. Were this not made it would require referencing of the cavity length change, $\Delta$, to the *nearest* transmission peak of a Gaussian beam, complicating both data collection and analysis. With these simplifications the relevant equation here becomes

$$\Delta = \frac{\lambda}{2}\frac{|l|}{\pi}arccos(\pm|1 - D/R|) = \frac{\lambda}{2}\frac{|l|}{\pi}arccos(D/R - 1). \quad (S15)$$

As a final step, the term on the right-most side of the equation emerges when the negative sign of the argument in brackets is taken. This selection is made in light of the fact that both *g*-parameters are negative and considering where on the stability diagram (Section A) the cavity in question sits.

The linear dependence on the resonant topological charge was further verified using the experimental data. Figure S2 shows this data (blue crosses) for the first four resonantly transmitted OAM states. The red line was a linear fit using the above equation performed using the curve fitting toolbox provided by MATLAB. The inferred cavity length using the slope of the line was found to be 70.60 microns, which agreed with our expectations based on the width of the components used to modify the FP cavity.

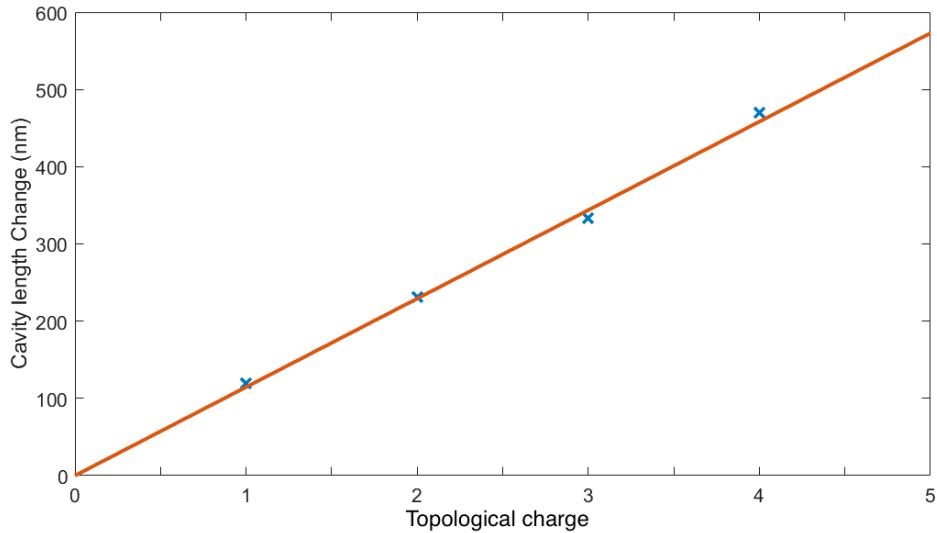

*Figure S2: The variation of the cavity length as a function of the resonant topological charge. Crosses represent experimental data, while the line is a fit using Equation (S15).*

### Section C—Details of experiments

Spatial light modulators (SLMs) are commonly used to produce beam of light carrying OAM. Here, an SLM (Boulder Nonlinear Systems, pixel pitch 15μm x 15μm, fill factor 83.4 %) was used to reflect an incident, linearly polarized beam from a collimated, frequency-stabilized HeNe laser (Thorlabs HRS015B, vacuum wavelength 632.991 nm). The output from the laser was sent through an optical isolator and spatially filtered using a 15-micron pinhole before being re-collimated to produce a clean Gaussian beam. The re-collimated beam was then reflected off the surface of the SLM to produce the desired OAM state. The SLM was set to display a phase-only interference pattern between a plane wave and a vortex beam of the desired topological charge.

The resulting beam was aligned to a scanning FP cavity (a modified Thorlabs SA200-5B, minimum finesse 250, original FSR 1.5 GHz, driven by an SA-201 piezo controller) with the output monitored by a CCD camera or a photodiode. Mode matching of a Gaussian beam to the cavity was performed using a thin lens of focal length 250 mm. Only small corrections were required to match the vortex beams to the cavity following this initial alignment. The transmitted signal was monitored via the output of a photodiode as a function of cavity length, which greatly facilitated fine mode-matching adjustments. Throughout this experiment, the modulation of the cavity length was sufficient to view greater than one full free spectral range (FSR) in each case, although individual spectra have been cropped for clarity. Figure S3 provides a schematic of the experimental setup.

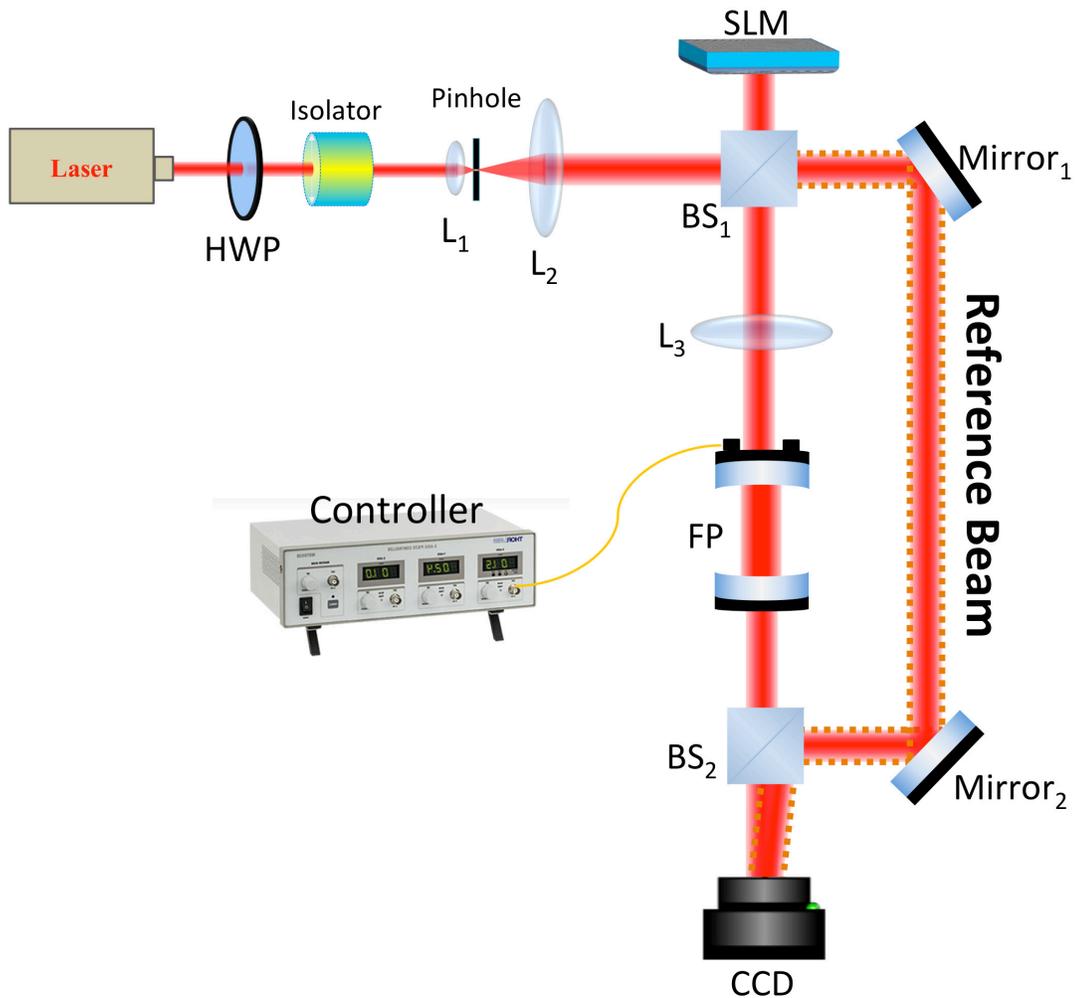

*Figure S3: Experimental setup. The linearly polarized beam of a stabilized HeNe laser is sent through a half-wave plate and optical isolator before being spatially filtered using a 15 μm pinhole. The half-wave plate was used to rotate the orientation of the (linear) polarization axis of*

*the beam. A telescope was incorporated into the spatial filtering of the beam (focal length of $L_1(L_2)$ was 11 mm(25.4 mm) to expand the spot size to facilitate mode matching to the cavity. The filtered beam then travelled through a 50/50 beam splitter ($BS_1$), which allowed normal incidence onto the SLM while also providing an additional reference Gaussian beam to form the forked interference patterns. The beam was reflected off of the SLM and returned through the beam splitter and mode-matched to the FP cavity using a thin lens of focal length 250 mm. The cavity length was controlled using the matched Thorlabs piezo controller, while the transmitted signal was monitored using either a CCD camera or photodiode.*

To measure the topological charge of the transmitted OAM state, a non-polarizing beam splitter prior to the SLM was used to sample part of the initial Gaussian beam. Two mirrors were used to redirect this beam to interfere with the transmitted beam at a slight angle at the CCD camera. An alternate piezo controller (Thorlabs MDT694B) was used to maintain the cavity length while imaging the transmitted beam.

It is important to highlight that the commercial scanning Fabry-perots (FPs) used throughout these experiments were modified from their off-the-shelf configuration. The Thorlabs SA200-5B FP cavities are equipped with high reflectivity (~99.375%) mirrors with 50 mm radii of curvature in a confocal configuration. This configuration results in a complete degeneracy of the longitudinal and transverse cavity modes, which transmits all resonant modes at identical cavity lengths. By increasing the cavity length by adding a small extension to one mirror of around 20 mm the degeneracy of the modified FP is broken, permitting differentiation of the transverse and longitudinal modes of the cavity while remaining in the stable cavity regime.

**Section D—Simulations of the resonant modes inside an FP cavity**

Numerical calculations were performed using FINESSE (Frequency domain Interferometer Simulation SoftwarE) (*5*) to confirm our experimental findings.

Figure S4 compares a FINESSE simulation and experimental data as a function of cavity length for different resonant OAM states. The two plots share the same legend for simplicity. The upper plot is the output of FINESSE while the lower plot is typical experimental data from a single cavity.

These numerical and experimental data present a qualitative match and serve to confirm our findings of the sensitivity of an FP cavity to OAM states.

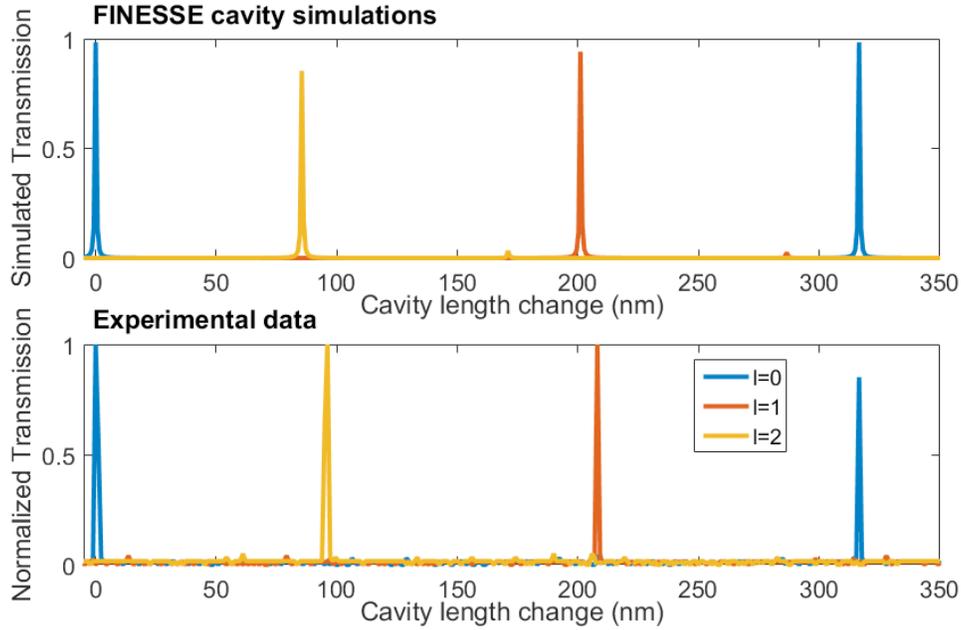

*Figure S4: (top) FINESSE simulations and (bottom) experimental data, as a function of the length variation of an FP cavity for three different OAM states.*

**Section E—Breaking the sign degeneracy of OAM states**

The main panel of Figure S5 shows the transmission of the scanning FP for two incident beams of a pure OAM state $|l=+1\rangle$ and of the superposition of $|l=+1\rangle \& |l=-1\rangle$, in blue and green, respectively. Clearly, the system is unable to differentiate the two incident signals carrying opposite OAM states as the blue and green transmission peaks occur at the identical cavity length. The red line shows the reference transmission of an incident beam of the superposition of $|l=0\rangle \& |l=+2\rangle$, launched via the appropriate display on the SLM. The cyan line, which shows a high level of similarity to the reference transmission, was initially a superposition of $|l=+1\rangle \& |l=-1\rangle$ that was sent through a spiral phase plate (SPP, n=+1), converting it to a superposition of $|l=0\rangle \& |l=+2\rangle$. The initially indistinguishable opposite OAM states can now be sorted in a fashion similar to other OAM states. Hence, a more intricate design consisting of two FP cavities as illustrated in Figure S6 can be used to replace the constituent module of an OAM sorter for breaking the sign degeneracy.

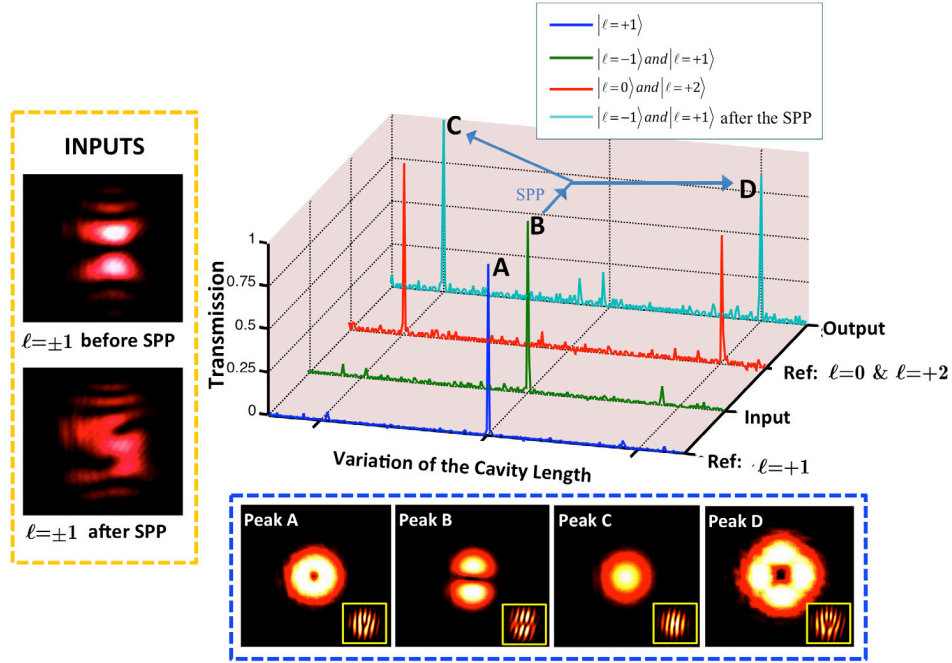

*Figure S5: Breaking the degeneracy of two OAM states with opposite signs. (main) Comparison of transmissions of single OAM state (l=+1, in blue), sign-degenerate OAM states (l=+1 & l=-1, in green), reference SLM-created superimposed states (l=0 & l=+2, in red) and the post-SPP degenerate pair (in cyan). The inset at the bottom shows the intensity distributions of the transmission peak labeled by letters A-D in the main figure. The panel inset on the left shows the intensity distribution of the degenerate pair before (top) and after (bottom) the SPP.*

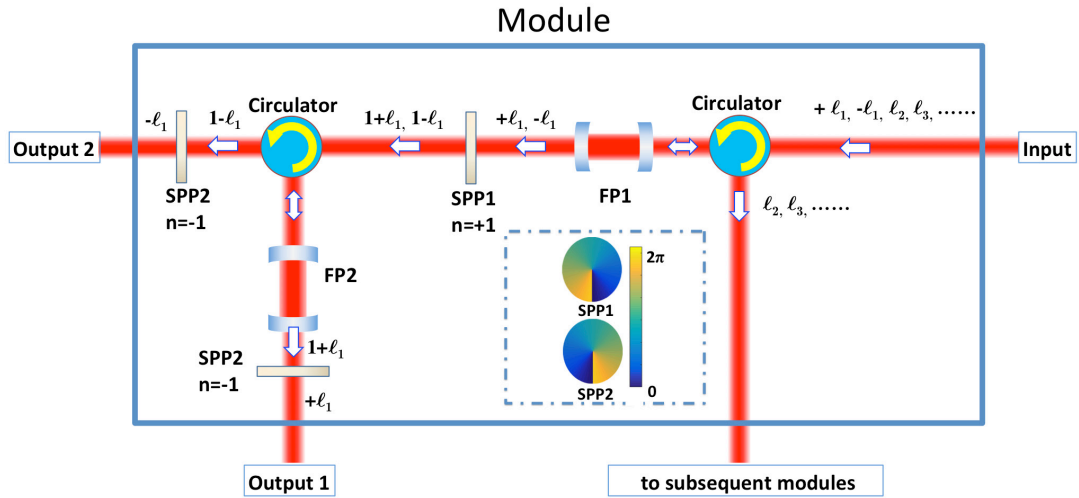

*Figure S6: Alternative design of the module that breaks the sign degeneracy of OAM states. Two output ports instead of one are now used for the topological charges with opposite signs.*

### Section F: Efficiency of the OAM sorting method

While FPs can be theoretically 100% efficient, in practice they are limited due to the surface roughness of the mirror coatings/surfaces, alignment errors and other imperfections. For a high finesse cavity such those used in cavity QED experiments, scattering and absorption losses from the mirrors are the major source of loss (*6*). A Thorlabs power meter was used to measure the incident cw laser power entering and exiting the cavity at resonance. A peak transmitted power of 21.7% was measured for a mode-matched Gaussian beam, comparable to the majority of published alternative OAM sorting methods that report absolute efficiencies.

While ~20 % transmission is acceptable in the lab, higher transmission is required for practical applications. Moving from the visible (~632.8 nm) to the infrared to minimize surface scattering and material absorption losses is the simplest method of improving efficiency. Optimization of the FP and mirrors would also increase efficiency, as would using dielectric mirrors. For example, an FP operating at a wavelength of 1064 nm has previously been reported with a transmission efficiency of 99.14 ± 0.86% with finesse above 2000 (*7*). Commercial etalons already used in Dense Wavelength Division Multiplexing devices also display efficiencies above 90 % with insertion losses below 1.5 dB at 1550 nm (*8*).

1. L. G. Gouy, *Sur une propriété nouvelle des ondes lumineuses*. (Gauthier-Villars, 1890).
2. H. Kogelnik, T. Li, Laser beams and resonators. *Applied optics* **5**, 1550-1567 (1966).
3. L. Allen, M. W. Beijersbergen, R. Spreeuw, J. Woerdman, Orbital angular momentum of light and the transformation of Laguerre-Gaussian laser modes. *Physical Review A* **45**, 8185 (1992).
4. A. E. Siegmann. (University Science Books, 1986).
5. A. Freise, D. Brown, C. Bond, Finesse, Frequency domain INterferomEter Simulation SoftwarE. *arXiv preprint arXiv:1306.2973*, (2013).
6. C. J. Hood, H. Kimble, J. Ye, Characterization of high-finesse mirrors: Loss, phase shifts, and mode structure in an optical cavity. *Physical Review A* **64**, 033804 (2001).
7. H. Sekiguchi *et al.*, Ultralow-loss mirror of the parts-in-10 6 level at 1064 nm. *Optics letters* **20**, 530-532 (1995).
8. Lightwaves2020, in *www.Lightwaves2020.com,* Lightwaves2020, Ed. (Lightwave2020, 2015).